\documentclass[journal]{IEEEtran}

\usepackage[utf8]{inputenc}
\usepackage{color}
\usepackage{xcolor}
\usepackage{array}
\usepackage{verbatim}
\usepackage{float}
\usepackage{amsmath}
\usepackage{amsthm}
\usepackage{amssymb}
\usepackage{graphicx}
\usepackage{longtable}
\usepackage{multirow}
\usepackage{booktabs}
\usepackage{longtable}

\usepackage[unicode=true,
bookmarks=false,
breaklinks=false,pdfborder={0 0 1},colorlinks=false]
{hyperref}
\hypersetup{
	colorlinks,bookmarksopen,bookmarksnumbered,citecolor=blue,urlcolor=blue}
\usepackage{cite}

\usepackage{lipsum}
\usepackage{mathtools}
\usepackage{cuted}

\usepackage{algorithmic}
\usepackage{longtable}

\floatstyle{ruled}
\newfloat{algorithm}{tbp}{loa}
\providecommand{\algorithmname}{Algorithm}
\floatname{algorithm}{\protect\algorithmname}

\makeatletter
\let\oldforeign@language\foreign@language
\DeclareRobustCommand{\foreign@language}[1]{%
	\lowercase{\oldforeign@language{#1}}}

\let\oldforeign@language\foreign@language
\DeclareRobustCommand{\foreign@language}[1]{%
	\lowercase{\oldforeign@language{#1}}}

\ifCLASSINFOpdf
\else
\fi

\@ifundefined{showcaptionsetup}{}{%
	\PassOptionsToPackage{caption=false}{subfig}}
\usepackage{subfig}

\usepackage{balance}

\ifCLASSINFOpdf
\else
\fi

\newtheorem{thm}{Theorem}
\newtheorem{rem}{Remark}

\newtheorem{assum}{Assumption}
\ifCLASSINFOpdf
\else
\fi

\def\ps@IEEEtitlepagestyle{%
	\def\@oddhead{\parbox[t][\height][t]{\textwidth}{\centering \scriptsize
			Personal use of this material is permitted. Permission from the author(s) and/or copyright holder(s), must be obtained for all other uses. Please contact us and provide details if you believe this document breaches copyrights.\\
			\noindent\makebox[\linewidth]{}
		}\hfil\hbox{}}%
	\def\@evenhead{\scriptsize\thepage \hfil \leftmark\mbox{}}%
	\def\@oddfoot{\parbox[t][\height][l]{\textwidth}{
			\vspace{-20pt}{\rule{\textwidth}{0.4pt}}\\ \footnotesize{\bf{\footnotesize\textcolor{red}{A. M. Ali, H. A. Hashim, and A. Jayasiri, "A Unified Finite-Time Sliding Mode Quaternion-based Tracking Control for Quadrotor UAVs without Time Scale Separation," The 2025 IEEE American Control Conference (ACC), Denver, Colorado, USA, 2025.}}}\\
			\noindent\makebox[\linewidth]
		}\hfil\hbox{}}%
	\def\@evenfoot{\MYfooter}}

\makeatother
\begin{document}
	\bstctlcite{IEEEexample:BSTcontrol}

\title{A Unified Finite-Time Sliding Mode Quaternion-based Tracking Control for Quadrotor UAVs without Time Scale Separation}

\author{Ali M. Ali, Hashim A. Hashim and Awantha Jayasiri
	\thanks{This work was supported in part by the National Sciences and Engineering Research Council of Canada (NSERC), under the grants RGPIN-2022-04937.}
	\thanks{Corresponding Author email: hhashim@carleton.ca - A. M. Ali and H. A. Hashim are with the Department of Mechanical
		and Aerospace Engineering, Carleton University, Ottawa, ON, K1S-5B6,
		Canada. A. Jayasiri is with the Flight Research Laboratory, National Research
		Council, Ottawa, ON K1B 1J8, Canada}
}



\maketitle
\begin{abstract}
This paper presents a novel design for finite-time position control
of quadrotor Unmanned Aerial Vehicles (UAVs). A robust, finite-time,
nonlinear feedback controller is introduced to reject bounded disturbances
in tracking tasks. The proposed control framework differs conceptually
from conventional controllers that utilize Euler angle parameterization
for attitude and adhere to the traditional hierarchical inner-outer
loop design. In standard approaches, the translational controller
and the corresponding desired attitude are computed first, followed
by the design of the attitude controller based on time-scale separation
between fast attitude and slow translational dynamics. In contrast,
the proposed control scheme is quaternion-based and utilizes a transit
feed-forward term in the attitude dynamics that anticipates the slower
translational subsystem. Robustness is achieved through the use of
continuously differentiable sliding manifolds. The proposed approach
guarantees semi-global finite-time stability, without requiring time-scale
separation. Finally, numerical simulation results are provided to
demonstrate the effectiveness of the proposed controller.
\end{abstract}

\section{Introduction}\label{introduction}

\IEEEPARstart{T}{racking} control of Quadrotor Unmanned Aerial Vehicles (UAVs) is a
challenging nonlinear feedback control problem, primarily due to the
nonlinear nature of the dynamics, underactuation, and outer disturbances
\cite{hash2025_RIENG_Avionics}. Over the past three decades, significant
contributions have addressed this complexity \cite{frazzoli2000trajectory,hashim2023exponentially,dierks2009output,lee2010geometric,lee2017geometric,ali2024_ASC_DRL_Dock,ali2024_ACC_MPC}.
In practical applications, disturbances such as wind must be mitigated
to achieve accurate pose (position and orientation) tracking. While
asymptotic stability ensures that error signals converge to zero as
time approaches infinity, finite-time stability is often more favorable
\cite{haddad2015finite}. Early work on nonlinear feedback control
for quadrotor UAVs includes techniques such as feedback linearization
\cite{mistler2001exact}, backstepping \cite{frazzoli2000trajectory},
and sliding mode control \cite{xu2006sliding}. The hierarchical inner-outer
loop design procedure has been widely adopted for the position control
of quadrotor UAVs \cite{frazzoli2000trajectory,shevidi2024adaptive,labbadi2020robust}.
These approaches are based on the time-scale separation principle
\cite{nascimento2019position}, which typically involves designing
a position controller first, followed by calculating the required
thrust and the desired vehicle attitude (orientation) to design the
attitude controller. In this framework, position and attitude control
are developed separately, requiring the inner loop (attitude subsystem)
to be asymptotically stable with a higher bandwidth than the outer
loop \cite{zagaris2004fast}.

Despite the popularity of inner-outer hierarchical schemes with time-scale
separation, these approaches have two key drawbacks: (i) high gains
for the inner-control loop relative to the outer-control loop are
required to maintain closed-loop stability, and (ii) in practice,
the computed control inputs (rotational torques and thrust) must be
mapped to the actual low-level controllers at the same frequency,
rather than at different frequencies. Designing the total thrust and
torques separately with different bandwidths is not recommended \cite{zuo2010trajectory}.
Most existing literature addresses the quadrotor UAV control problem
using Euler angle parameterization for attitude control \cite{wang2023robust,labbadi2020robust}.
However, this approach can lead to kinematic singularities (known
as gimbal lock), which may cause control failures in certain configurations.
To address both singularity and time-scale separation issues, a Lyapunov-based
control design on $\mathbb{S\mathbb{E}}(3)$ was introduced in \cite{lee2010geometric}
and later modified to reject unstructured disturbances \cite{goodarzi2015geometric,bisheban2020geometric}.
Although these results ensure asymptotic stability, they do not guarantee
finite-time convergence.

\paragraph*{Contributions}

Accordingly, there is a need for robust controllers that are free
from singularities, ensure finite-time stability, and do not rely
on the separation of time scales between translational and attitude
dynamics. Motivated by these challenges, this paper proposes a unified
finite-time sliding mode control for quadrotor UAVs that guarantees
finite-time convergence without time-scale separation between position
and attitude control, unlike the approaches in \cite{shevidi2024adaptive,labbadi2020robust}.
The proposed attitude controller serves as the primary mechanism,
stabilizing both translational and attitude dynamics through a unified
Lyapunov function while rejecting unknown bounded wind disturbances.
The proposed approach is quaternion-based to avoid kinematic and attitude
singularities while ensuring finite-time convergence, in contrast
to \cite{kang2020quaternion,kumar2020quaternion}. Additionally, the
control strategy guarantees semi-global finite-time stability of the
closed-loop error signals.

\paragraph*{Organization}

The remainder of the paper is organized as follows: Section \ref{sec:Preliminaries}
presents the preliminaries and mathematical notation. The problem
formulation is outlined in Section \ref{sec:Problem Formulation}.
Section \ref{sec:Control Synthesis} introduces the proposed control
scheme along with the corresponding stability analysis. The effectiveness
of the proposed approach is demonstrated through numerical simulations
in Section \ref{sec: Numerical Results}. Finally, Section \ref{sec: Conclusion}
provides concluding remarks.

\section{Preliminaries and Math Notations \label{sec:Preliminaries}}

In this paper, the set of non-negative real numbers, real $a-$dimensional
space, and real $a\times b$ dimensional space are referred to as
$\mathbb{R}_{+}$, $\mathbb{R}^{a}$, and $\mathbb{R}^{a\times b}$,
respectively. The $a$-by-$a$ identity matrix is defined as $\mathbb{I}_{a}$.
For all $x\in\mathbb{R}^{a}$, $^{\top}$ denotes a transpose of $x$
and $\Vert x\Vert=\sqrt{x^{\top}x}$ stands for the Euclidean norm
of $x\in\mathbb{R}^{a}$. The $a$-sphere $\mathcal{\mathrm{S}}^{a}$
is defined as $\mathcal{\mathrm{S}}^{a}=\{x\in\mathbb{R}^{a+1}\ |\,\Vert x\Vert=1\}.$
$R\in\mathbb{S\mathbb{O}}(3)$ describes the unique representation
of the attitude $\mathbb{S\mathbb{O}}(3)=\{R\in\mathbb{R}^{3\times3}\ |\,R^{\top}R=\mathbb{I}_{3},det(R)=+1\}$
of a rigid body in the body-frame $\{\mathcal{B}\}$ relative to the
inertial-frame $\{\mathcal{I}\}$ in 3D space with the three orthonormal
basis $(x_{I},y_{I},z_{I})$ being $x_{I}=[\begin{array}{ccc}
	1 & 0 & 0\end{array}]^{\top}$, $y_{I}=[\begin{array}{ccc}
	0 & 1 & 0\end{array}]^{\top}$, and $z_{I}=[\begin{array}{ccc}
	0 & 0 & 1\end{array}]^{\top}.$ The Lie-algebra of the group $\mathbb{S\mathbb{O}}(3)$ is defined
by $\mathfrak{so}(3)$ and expressed as \cite{hash2019_arXiv_Special_Survey, hashim2021gps, ghanizadegan2025deepukf}
\[
\mathfrak{so}(3)=\{[x]_{\times}\in\mathbb{R}^{3\times3}\ |\ [x]_{\times}^{\top}=-[x]_{\times}\},
\]
with $[x]_{\times}$ standing for a skew-symmetric matrix such that
the map $[\cdot]_{\times}:\mathbb{R}^{3}\rightarrow\mathfrak{so}(3)$
is given by{\small{}
	\[
	[x]_{\times}=\left[\begin{array}{ccc}
		0 & -x_{3} & x_{2}\\
		x_{3} & 0 & -x_{1}\\
		-x_{2} & x_{1} & 0
	\end{array}\right]\in\mathfrak{so}(3),\hspace{1em}x=\left[\begin{array}{c}
		x_{1}\\
		x_{2}\\
		x_{3}
	\end{array}\right]
	\]
}The unit-quaternion is a non-Euclidean global non singular four-element
parametrization of the attitude $R\in\mathbb{S\mathbb{O}}(3)$ and
is denoted by
\[
Q=[\begin{array}{cccc}
	q_{1} & q_{2} & q_{3} & q_{0}\end{array}]^{\top}=[q^{\top},q_{0}]^{\top}\in\mathcal{\mathrm{S}}^{3}
\]
where $q=[q_{1},q_{2},q_{3}]^{\top}\in\mathbb{R}^{3}$ and $q_{0}\in\mathbb{R}$.
The multiplication operator of unit-quaternion vectors is denoted
by $\odot$ such that
\begin{alignat*}{1}
	Q_{3} & =Q_{1}\odot Q_{2}=\left[\begin{array}{c}
		q_{1}\\
		q_{01}
	\end{array}\right]\odot\left[\begin{array}{c}
		q_{2}\\
		q_{02}
	\end{array}\right]\\
	& =\left[\begin{array}{c}
		q_{01}q_{2}+q_{02}q_{1}+[q_{1}]_{\times}q_{2}\\
		q_{01}q_{02}-q_{1}^{\top}q_{2}
	\end{array}\right]
\end{alignat*}
The inverse and neutral element of $Q$ is defined by $Q^{-1}=[-q^{\top},q_{0}]^{\top}$
and $Q_{I}=[0_{1\times3},1]^{\top}$, respectively. The rotation matrix
$R(Q)\in\mathbb{S\mathbb{O}}(3)$ described with respect to $Q$ is
given by \cite{hash2019_arXiv_Special_Survey}:
\begin{equation}
	R(Q)=(q_{0}^{2}-qq^{\top})\mathbb{I}_{3}+2qq^{\top}-2q_{0}[q]_{\times}\label{eq:rodriguez}
\end{equation}
Let $Q_{d}=\left[q_{d}^{\top},q_{d0}\right]^{\top}\in\mathcal{\mathrm{S}}^{3}$
be the desired unit-quaternion vector, where $q_{d}\in\mathbb{R}^{3}$
and $q_{d0}\in\mathbb{R}$. Consider the following map $\Xi(\cdot):\mathbb{R}^{4}\rightarrow\mathbb{R}^{4\times4}$:
\begin{equation}
	\Xi(Q)=\frac{1}{2}\left[\begin{array}{c}
		q_{0}\mathbb{I}_{3}+[q]_{\times}\\
		-q^{\top}
	\end{array}\right]\label{eq:kinematics-matrix}
\end{equation}
Let $\text{sgn}(x):\mathbb{R}\rightarrow\{-1,0,1\}$ to be a discontinuous
function given by
\[
\text{sgn}(x)=\begin{cases}
	-1 & \text{if}\hspace{0.2cm}x<0,\\
	0 & \text{if}\hspace{0.2cm}x=0,\\
	1 & \text{if}\hspace{0.2cm}x>0.
\end{cases}
\]
Define $\overline{q}=\left[\begin{array}{ccc}
	\tilde{q}_{2} & -\tilde{q}_{1} & -\tilde{q_{0}}\end{array}\right]^{\top}\in\mathbb{R}^{3}$, the identities below will be used in the subsequent derivations:
\begin{alignat}{1}
	(R(Q)^{\top}-R(Q_{d})^{\top})z_{I} & =2R(Q)^{\top}[\overline{q}]_{\times}\tilde{q}\label{eq:17}
\end{alignat}

\section{Problem Formulation \label{sec:Problem Formulation}}

\begin{figure}[h]
	\includegraphics[width=9cm,height=7cm,keepaspectratio]{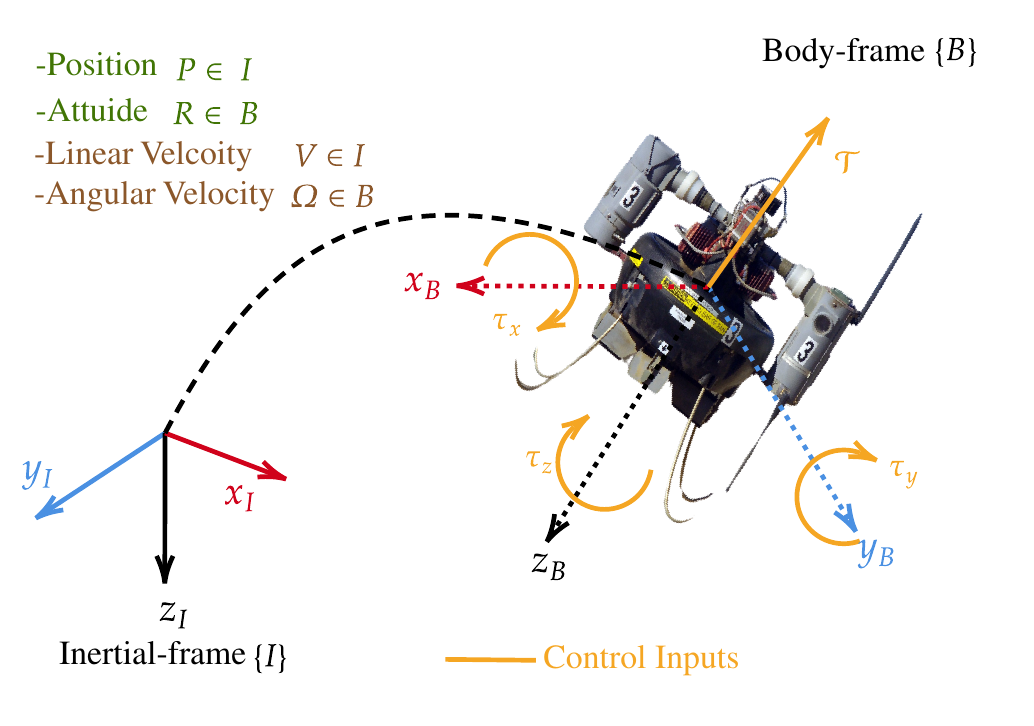}
	
	\caption{Position tracking problem of an under-actuated quadrotor UAV in 3D
		space.}
	\label{fig:1}
\end{figure}

Let $P=[p_{x},p_{y},p_{z}]^{\top}\in\mathbb{R}^{3}$ and $V=[v_{x},v_{y},v_{z}]^{\top}\in\mathbb{R}^{3}$
denote the position and the linear velocity of the quadrotor UAV,
respectively, with $P,V\in\{\mathcal{I}\}$. The angular velocity
is denoted by $\Omega=[\omega_{x},\omega_{y},\omega_{z}]^{\top}\in\mathbb{R}^{3}$.
The dynamics of a quadrotor UAV are described as follows \cite{hashim2023exponentially}:
\begin{flalign}
	\text{Translation} & \begin{cases}
		\dot{P} & =V\\
		\dot{V} & =-\frac{\mathcal{T}}{m}R(Q)^{\top}z_{I}+gz_{I}+\frac{F_{d}}{m}
	\end{cases}\label{eq:31}\\
	\text{Attitude} & \begin{cases}
		\dot{Q} & =\frac{1}{2}\Xi(Q)\Omega\\
		J\dot{\Omega} & =-[\Omega]_{\times}J\Omega+\Gamma
	\end{cases},\hspace{0.25cm}Q,\Omega\in\{\mathcal{B}\}\label{eq:32}
\end{flalign}
with $\Xi(Q)$ being defined in \eqref{eq:kinematics-matrix}, $J\in\mathbb{R}^{3\times3}$
being the inertia matrix, $\mathcal{T\in\mathbb{R}}$ being the total
propellers thrust acting on the direction $z_{I}$ (see \ref{fig:1}).
$\Gamma=\left[\begin{array}{ccc}
	\tau_{x} & \tau_{y} & \tau_{z}\end{array}\right]^{\top}\in\{\mathcal{B}\}$ is the torque control acting on each axis. $F_{\text{d}}$ is an
unknown bounded force caused by wind. Our main objective is to design
control laws for $\mathcal{T}$ and $\Gamma$ to track a time-varying
reference position $P_{d}.$

\begin{assum}\label{assum:Model assumpations}For the quadrotor UAV
	model shown in \eqref{eq:31}, \eqref{eq:32}, and described in Fig.
	\ref{fig:1} the following assumptions hold:
	\begin{itemize}
		\item[A1.] \label{A1.The-VTOL-UAV-model}The quadrotor UAV model has a fixed
		mass $m$ and the inertia matrix $J$ is a constant symmetric positive
		definite matrix.
		\item[A2.] \label{A2.The-VTOL-UAV-current}The desired position $P_{d}$ and
		its derivatives $\dot{P_{d}}=\dot{V_{d}}$, $\ddot{P_{d}}$, $\dddot{P_{d}}$,
		and $\ddddot{P}_{d}$ are assumed to be smooth and uniformly upper
		bounded.
		\item[A3.] \label{A2.The-VTOL-UAV-dist}The disturbance force $F_{d}$ is upper
		bounded by $\delta_{f}=\left[\begin{array}{ccc}
			\delta_{fx} & \delta_{fx} & \delta_{fx}\end{array}\right]^{\top}\in\mathbb{R}^{3}$, such that $F_{d}\in\Delta_{f}:=\{F_{d}\in\mathbb{R}^{3},||F_{d}||<||\delta_{f}||\}$.
	\end{itemize}
\end{assum}

The rotational dynamics in \eqref{eq:32} are fully actuated, as the
torque control inputs, $\Gamma$, directly influence all angular accelerations,
and consequently the angular velocities and attitude. In contrast,
the translational dynamics are more challenging to control, as there
is only a single input, $\mathcal{T}$, acting in the direction of
$z_{\text{B}}$, which can only influence vertical acceleration. Referring
to \eqref{eq:31}, by adding and subtracting $\frac{\mathcal{T}}{m}R(Q_{d})^{\top}z_{B}$
to $\dot{V},$ where $Q_{d}$ is the required unit-quaternion to be
defined later in the attitude control design, $\dot{V}$ can be rewritten
as follows:
\begin{equation}
	\dot{V}=\mathcal{F}-\frac{\mathcal{T}}{m}(R(Q)^{\top}-R(Q_{d})^{\top})z_{I}+\frac{F_{d}}{m}\label{eq:new-dynamics}
\end{equation}
with
\begin{equation}
	\mathcal{F}=gz_{I}-\frac{\mathcal{T}}{m}R(Q_{d})^{\top}z_{I}\label{eq:virtual_controller}
\end{equation}
where $\mathcal{F}=\left[\begin{array}{ccc}
	f_{x} & f_{y} & f_{z}\end{array}\right]^{\top}\in\mathbb{R}^{3}$ denotes a virtual control input that has full control on the translational
acceleration.

\subsection{Errors and Error Dynamics}

The error between the desired and the true unit-quaternion is expressed
as:
\begin{alignat}{1}
	\tilde{Q} & =Q_{d}^{-1}\odot Q=\left[\begin{array}{c}
		q_{d0}q_{0}-q_{0}q_{d}-[q_{d}]_{\times}q\\
		q_{d0}q_{0}+q_{d}^{\top}q
	\end{array}\right]\label{eq:quaterion_error}
\end{alignat}
where $\tilde{Q}=[\tilde{q}^{\top},\tilde{q}_{0}]^{\top}\in\mathcal{\mathrm{S}}^{3}$
for all $\tilde{q}_{0}\in\mathbb{R}$ and $\tilde{q}\in\mathbb{R}^{3}$.
Define $\tilde{P}=P-P_{d}$, $\tilde{V}=V-V_{d}$, $\tilde{\Omega}=\Omega-\Omega_{d},$
and $\dot{\tilde{V}}=\dot{V}-\dot{V}_{d}$ as the position, linear
velocity, angular velocity, and linear acceleration error signals.
In contrast to \cite{shevidi2024adaptive} and \cite{labbadi2020robust},
in this paper we design the virtual control input $\mathcal{F}$ to
drive $\dot{V}\rightarrow\dot{V_{d}}$ overcoming the disturbance
$\frac{F_{d}}{m}$, while $\frac{\mathcal{T}}{m}(R(Q)^{\top}-R(Q_{d})^{\top})z_{I}$
can be left as perturbation to the translational  acceleration. While
the rotational torque input $\Gamma$ is driving the attitude $Q\rightarrow Q_{d}$,
the perturbation term $\frac{\mathcal{T}}{m}(R(Q)^{\top}-R(Q_{d})^{\top})z_{I}\rightarrow0$.
Let us introduce an auxiliary variable $\Theta=[\theta_{1},\theta_{2},\theta_{3}]^{\top}\in\mathbb{R}^{3}$
such that
\begin{equation}
	\Theta=\tilde{\Omega}-\Psi\label{eq:auxiliary_variable}
\end{equation}
where $\Psi\in\mathbb{R}^{3}$ is a design parameter to be defined
subsequently. The attitude error dynamics $\dot{\tilde{Q}}$ can be
described as follows:
\begin{equation}
	\dot{\tilde{Q}}=\frac{1}{2}\Xi(\tilde{Q})\tilde{\Omega},\hspace{1em}\Xi(\tilde{Q})=\frac{1}{2}\left[\begin{array}{c}
		\tilde{q}_{0}\mathbb{I}_{3}+[\tilde{q}]_{\times}\\
		-\tilde{q}^{\top}
	\end{array}\right]\label{eq:error_kinematics}
\end{equation}

\begin{rem}
	The selection of $\Psi$ plays a crucial role in ensuring the closed-loop
	stability of the interconnected translational and attitude systems
	in \eqref{eq:31} and \eqref{eq:32}. In control schemes that rely
	on time-scale separation, $\tilde{\Omega}\rightarrow0$ faster than
	$\tilde{V}\rightarrow0$ to guarantee closed-loop stability. In the
	proposed work, our objective is to first drive $\tilde{\Omega}$ to
	an intermediate variable $\Psi$, and subsequently drive $\Psi$ to
	zero. The design of $\Psi$ should be directly proportional to the
	linear velocity tracking error $\tilde{V}$, ensuring that $\tilde{\Omega}\rightarrow\Psi(\tilde{V})$
	as $\Theta\rightarrow0$, which ultimately leads to $\tilde{\Omega}\rightarrow0$
	as $\Psi(\tilde{V})\rightarrow0$. This facilitates the stabilization
	of both attitude and translational dynamics without relying on the
	time-scale separation. 
\end{rem}
After designing the virtual control $\mathcal{F}$ in \eqref{eq:virtual_controller},
the desired attitude $Q_{d}$ must be computed to track the virtual
controller using the attitude controller. However, solving \eqref{eq:virtual_controller}
directly results in an underdetermined system, as there are three
known values (the components of the virtual controller) and four unknowns
in $Q_{d}$ leading to infinitely many solutions. One possible approach
to obtain the desired attitude $Q_{d}$ and the total thrust $\mathcal{T}$
is as follows:
\begin{equation}
	\mathcal{T}=m\Vert gz_{I}-\mathcal{F\Vert}\label{eq:total_thrust}
\end{equation}
\begin{equation}
	q_{d}=\left[\begin{array}{c}
		\frac{mf_{y}}{2\mathcal{T}q_{d0}}\\
		\frac{-mf_{x}}{2\mathcal{T}q_{d0}}\\
		0
	\end{array}\right],\hspace{1em}q_{d0}=\sqrt{\frac{m}{2\mathcal{T}}(g-f_{z})+\frac{1}{2}}\label{eq:Qd}
\end{equation}
Furthermore, the desired rotational velocity can be written as follows:
\begin{equation}
	\dot{\Omega}_{d}=\Phi(\mathcal{F})\dot{\mathcal{F}}\label{eq:desired-rate}
\end{equation}
with{\small{}
	\begin{equation}
		\Phi(\mathcal{F})=\frac{1}{\alpha_{1}^{2}\alpha_{2}}\left[\begin{array}{ccc}
			-f_{x}f_{y} & -f_{y}^{2}+\alpha_{1}\alpha_{2} & f_{y}\alpha_{2}\\
			f_{x}^{2}-\alpha_{1}\alpha_{2} & f_{x}f_{y} & -f_{1}\alpha_{2}\\
			f_{y}\alpha_{1} & -f_{1}\alpha_{1} & 0
		\end{array}\right]\label{eq:desired}
	\end{equation}
}where $\alpha_{1}=\Vert gz_{I}-\mathcal{F}\Vert$, and $\alpha_{2}=\Vert gz_{I}-\mathcal{F}\Vert+g-f_{z}.$
It is important to highlight that, if the desired acceleration is
provided such that $\dot{V}_{d}\neq\left[0,0,x\right]^{\top},$ for
all $x\geq g$, the extraction of the desired attitude $Q_{d}$ in
\eqref{eq:Qd} is free from singularities, for more information consult
\cite{hashim2023observer}. Furthermore the desired rotational acceleration
$\dot{\Omega}_{d}$ can be obtained as $\dot{\Omega}_{d}=\dot{\Phi}(\mathcal{F},\dot{\mathcal{F}})+\Phi(\mathcal{F})\ddot{\mathcal{F}}$
with $\dot{\Phi}(\mathcal{F},\dot{\mathcal{F}})$ being the time derivative
of \eqref{eq:desired}. 

\section{Control Synthesis \label{sec:Control Synthesis}}

In this section, the translational and attitude error dynamics are
formulated and the virtual translational  controller is synthesized.
Next, the attitude control is designed. Finally, the stability analysis
of the proposed control laws are discussed.

\subsection{Control design for translation dynamics \label{subsec:error_dynamics}}

\begin{figure}[h]
	\centering\includegraphics[width=8.7cm,height=7cm]{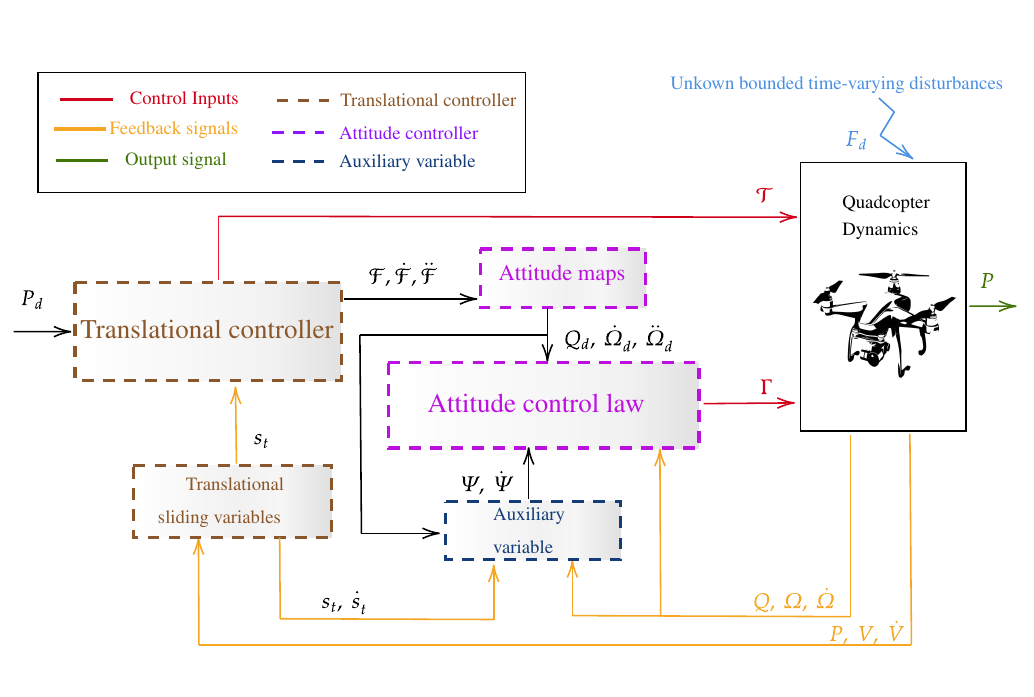}
	
	\caption{\textcolor{black}{An illustrative diagram of the proposed tracking
			control for quadrotor UAV. The sliding variables and control law are
			described in \eqref{eq:s_x}-\eqref{eq:s_z}. The virtual translational
			controller $\mathcal{F}$, along with its time derivative (defined
			in \eqref{eq:F_dot} and \eqref{eq:F_dot_dot}), is used to compute
			the desired attitude $Q_{d}$, as well as the desired angular velocity
			$\dot{\Omega}_{d}$ and acceleration $\ddot{\Omega}_{d}$, based on
			the attitude maps presented in \eqref{eq:Qd}, \eqref{eq:desired-rate},
			and \eqref{eq:desired}. The attitude control law generates the torque
			input by utilizing the translational sliding variables and the auxiliary
			variable, as defined in \eqref{eq:desgin_aux}, and \eqref{eq:Psi_dot},
			respectively.}}
	
	\label{Fig:2}
\end{figure}

Based on the earlier definitions of $\tilde{P}$, $\tilde{V}$, $\tilde{Q}$,
and $\tilde{\Omega}$, the translational and attitude dynamics presented
in \eqref{eq:31} and \eqref{eq:32}, along with the identity in \eqref{eq:17},
the error dynamics can be written as follows:
\begin{flalign}
	\dot{\tilde{P}}= & \tilde{V}\label{eq:33}\\
	\dot{\tilde{V}}= & \frac{-2\mathcal{T}}{m}R(Q)^{\top}[\overline{q}]_{\times}\tilde{q}+\mathcal{F}-\dot{V}_{d}+\frac{F_{d}}{m}\label{eq:34}\\
	\dot{\tilde{Q}}= & \frac{1}{2}\Xi(\tilde{Q})\tilde{\Omega}\label{eq:35}\\
	J\dot{\tilde{\Omega}}= & \Gamma-[\Omega]_{\times}J\Omega+J[\tilde{\Omega}]_{\times}R(\tilde{Q})\Omega_{d}-JR(\tilde{Q})\dot{\Omega}_{d}\label{eq:36}
\end{flalign}
Recall the translational error dynamics \eqref{eq:33} and \eqref{eq:34},
our objective is to design $\mathcal{F}$ to overcome the unknown
bounded disturbances. Let us define the following sliding variable
$s_{t}=\left[\begin{array}{ccc}
	s_{1} & s_{2} & s_{3}\end{array}\right]^{\top}\in\mathbb{R}^{3}$ defined as follows:
\begin{alignat}{1}
	s_{1} & =k_{s_{t}}|\tilde{p}_{x}|^{\beta_{_{t}}}\text{sgn}(\tilde{p}_{x})+\tilde{v}_{x}\label{eq:s_x}\\
	s_{2} & =k_{s_{t}}|\tilde{p}_{y}|^{\beta_{_{t}}}\text{sgn}(\tilde{p}_{y})+\tilde{v}_{y}\label{eq:s_y}\\
	s_{3} & =k_{s_{t}}|\tilde{p}_{z}|^{\beta_{_{t}}}\text{sgn}(\tilde{p}_{z})+\tilde{v}_{z}\label{eq:s_z}
\end{alignat}
with $k_{s_{t}}$ is a positive scalar gain and $\beta_{t}\in(0,1)$.
The virtual controller $\mathcal{F}$ ultimate goal is to stir the
error dynamics to the manifolds $s_{1}=k_{s_{t}}|\tilde{p}_{x}|^{\beta_{_{t}}}\text{sgn}(\tilde{p}_{x})+\tilde{v}_{x}=0,$
$s_{2}=k_{s_{t}}|\tilde{p}_{y}|^{\beta_{_{t}}}\text{sgn}(\tilde{p}_{y})+\tilde{v}_{y}=0$,
and $s_{3}=k_{s_{t}}|\tilde{p}_{z}|^{\beta_{_{t}}}\text{sgn}(\tilde{p}_{z})+\tilde{v}_{z}=0$.
On those manifolds, the dynamics will be governed by $\dot{\tilde{p}}_{x}=k_{s_{t}}|\tilde{p}_{x}|^{\beta_{t}}\text{sgn}(\tilde{p}_{x})$,
$\dot{\tilde{p}}_{y}=k_{s_{t}}|\tilde{p}_{y}|^{\beta_{t}}\text{sgn}(\tilde{p}_{y}),$
and $\dot{\tilde{p}}_{z}=k_{s_{t}}|\tilde{p}_{z}|^{\beta_{t}}\text{sgn}(\tilde{p}_{z})$.
Choosing $k_{s_{t}}>0$ and $\beta_{t}\in(0,1)$ guarantees that $\tilde{P}\rightarrow0.$
At this point, the previously mentioned manifolds become invariant
manifolds, such that the error dynamics are trapped and converge to
zero in a finite time. In fact, $\mathcal{F}$ is used to extract
the desired attitude, therefore it should be at least twice differentiable
to compute the desired attitude rate and acceleration in a closed
form. Now let us introduce the proposed virtual control input $\mathcal{F}$
as follows:
\begin{equation}
	\mathcal{F}=\dot{V}_{d}-k_{t}\mathcal{S}(s_{t})+\Upsilon\label{eq:virtual_controller_main}
\end{equation}
where $k_{t}=\left[\begin{array}{ccc}
	k_{1} & k_{2} & k_{3}\end{array}\right]^{\top}\in\mathbb{R}^{3}$ with $k_{1},$ $k_{2},$ and $k_{3}$ are positive scalar gain. $\mathcal{S}(s_{t}):\mathbb{R}^{3}\rightarrow\mathbb{R}^{3}$
is saturation function given as follows: 
\begin{equation}
	\mathcal{S}(s_{t})=\left[\begin{array}{c}
		\text{tanh}(s_{1})\\
		\text{tanh}(s_{2})\\
		\text{tanh}(s_{3})
	\end{array}\right]\label{eq:sat_function}
\end{equation}
where $\Upsilon\in\mathbb{R}^{3}$ is a design parameter to be proposed
later. Note that we have chosen the elements of $\mathcal{S}(\cdotp)$
to be continuously differentiable smooth functions everywhere expect
the origin.
\begin{rem}
	The sliding variables introduced in \eqref{eq:s_x}-\eqref{eq:s_z},
	are selected to ensure the continuity and differentiability of the
	virtual control law in \eqref{eq:virtual_controller_main}. The differentiability
	of $\mathcal{F}$ is a critical requirement for achieving the desired
	attitude, attitude rate, and attitude acceleration (visit \eqref{eq:Qd}
	and \eqref{eq:desired}). 
\end{rem}
The first and second time derivative of $s_{t}$ defined in \eqref{eq:s_x}-\eqref{eq:s_z}
are given by:
\begin{alignat}{1}
	\dot{s}_{t} & =\beta_{t}k_{s_{t}}\left(|\tilde{P}|^{\beta_{t}-1}\dot{\tilde{P}}\right)+\dot{\tilde{V}}\label{eq:s_t_dot}\\
	\ddot{s}_{t} & =k_{s_{t}}\beta_{t}\left(\left(\beta_{t}-1\right)|\tilde{P}|^{\beta_{t}-2}\text{sgn}(\dot{\tilde{P}})\dot{\tilde{P}}^{2}+|\text{\ensuremath{\tilde{P}}}|^{\beta_{t}-1}\ddot{\tilde{P}}\right)+\ddot{\tilde{V}}\label{eq:s_t_dot_dot}
\end{alignat}
Note that multiplication in $\dot{s}_{t}$ and $\ddot{s}_{t}$ is
elementwise. The first and second time derivative of $\mathcal{F}$
can be written as follows:
\begin{alignat}{1}
	\dot{\mathcal{F}} & =\ddot{V}_{d}-k_{t}\mathcal{D}(s_{t})\dot{s}_{t}+\dot{\Upsilon}\label{eq:F_dot}\\
	\ddot{\mathcal{F}} & =\text{\ensuremath{\dddot{V}_{d}-k_{t}\dot{\mathcal{D}}(s_{t})\dot{s}_{t}-k_{t}\mathcal{D}(s_{t})\ddot{s}_{t}}+\ensuremath{\ddot{\Upsilon}}}\label{eq:F_dot_dot}
\end{alignat}
with $\mathcal{D}(s_{t})=\text{diag\ensuremath{\left[\frac{\partial\text{tanh}(s_{i})}{\partial s_{i}}\right]}, with \ensuremath{i=1,2,3,} and \ensuremath{\dot{\mathcal{D}}(s_{t})}}$
is the time derivative of $\mathcal{D}(s_{t})$. One obtains $\ddot{\tilde{V}}$
as follows:
\begin{alignat}{1}
	\ddot{\tilde{V}} & =\dot{\mathcal{F}}-\ddot{V}_{d}+\frac{(gz_{I}-\mathcal{F})^{\top}\mathcal{\dot{F}}}{||gz_{I}-\mathcal{F}||}R(Q)^{\top}[\bar{q}]_{\times}\tilde{q}\nonumber \\
	& -\frac{2\mathcal{T}}{m}R(Q)^{\top}[\dot{\bar{q}}]_{\times}\tilde{q}-\frac{2\mathcal{T}}{m}R(Q)^{\top}[\bar{q}]_{\times}\dot{\tilde{q}}\label{eq:V_delta_dot_dot}
\end{alignat}
The fractional power in sliding variables \eqref{eq:s_x}-\eqref{eq:s_z}
is an instrumental tool for the finite-time convergence. However,
it introduces singular terms in the time derivative of the sliding
variables. In particular, $\dot{s}_{t}$ and $\ddot{s}_{t}\rightarrow\infty$
when $\tilde{P}=0\mid\dot{\tilde{P}}\neq0$ (visit \eqref{eq:s_t_dot}
and \eqref{eq:s_t_dot_dot}). To solve the singularity problem of
the sliding variables, let us define boundary layers around the vertical
axes when $\tilde{p}_{i}=0\mid\dot{\tilde{p}}_{i}\neq0$ as follows
$\mathcal{N}_{i}:=\left\{ (\tilde{p}_{i},\tilde{v}_{i})\in\mathbb{R}^{2}:\tilde{p}_{i}\in(-\epsilon,\epsilon)\right\} ,\forall i=\{x,y,z\},$
where $\epsilon<0.$ As such, we switch $\beta_{t}$ to be equal to
one when $(\tilde{p}_{i},\tilde{v}_{i})\in\mathcal{N}_{i},$ $\forall i=\{x,y,z\}$
and $\beta_{t}\in(0,1)$ otherwise. The objective of the attitude
control is (i) to track the time-varying desired attitude $Q_{d}$
obtained from the attitude maps in \eqref{eq:Qd}, \eqref{eq:desired-rate},
and \eqref{eq:desired}, (ii) to ensure the closed-loop stability
of both attitude and translational dynamics. In the light of the earlier
definition of the auxiliary variable $\Theta$ in \eqref{eq:auxiliary_variable},
the design variable $\Psi$ is designed as follows:
\begin{equation}
	\Psi=-k_{\eta}\tilde{q}+\frac{2\mathcal{T}}{m}[\overline{q}]_{\times}^{\top}R(Q)s_{t}\label{eq:desgin_aux}
\end{equation}
The attitude controller $\Gamma$ is designed as follows:
\begin{alignat}{1}
	\Gamma= & [\Omega]_{\times}J\Omega-J[\tilde{\Omega}]_{\times}R(\tilde{Q})\Omega_{d}+JR(\tilde{Q})\dot{\Omega}_{d}\nonumber \\
	- & k_{\theta}\Theta-k_{q}\tilde{q}+J\dot{\Psi}\label{eq:control_troque-1}
\end{alignat}
where $k_{\theta},k_{q}>0$ and $\dot{\Psi}$ is given by
\begin{alignat}{1}
	\dot{\Psi} & =-k_{\eta}(\dot{\tilde{q}})+\frac{2m}{\mathcal{T}}(gz_{I}-\mathcal{F})^{\top}\dot{\mathcal{F}}[\overline{q}]_{\times}^{\top}R(Q)s_{t}\nonumber \\
	& +\frac{2\mathcal{T}}{m}\left([\overline{q}]_{\times}^{\top}R(Q)\dot{s}_{t}+[\dot{q}]_{\times}^{\top}R(Q)s_{t}-[\overline{q}]_{\times}^{\top}R[\Omega]_{\times}s_{t}\right)\label{eq:Psi_dot}
\end{alignat}
Note that the power raised in \eqref{eq:Psi_dot} is elementwise.
Fig. \eqref{Fig:2} illustrates the proposed control scheme, where
the torque control input serves as the primary controller receiving
feedback measurements for both translational and attitude.

\subsection{Overall stability analysis\label{subsec:stability}}

In this section, the closed-loop stability will be analyzed considering
the total thrust, virtual position control, and attitude laws proposed
in \eqref{eq:total_thrust}, \eqref{eq:virtual_controller_main},
and \eqref{eq:control_troque-1}, respectively. 
\begin{thm}
	Consider the quadrotor UAV dynamics described in \eqref{eq:31} and
	\eqref{eq:32} with the control laws in \eqref{eq:total_thrust},
	\eqref{eq:virtual_controller_main}, and \eqref{eq:control_troque-1}
	with $\Upsilon$ is defined as follows
	\begin{equation}
		\Upsilon=-\beta\mathcal{S}(s_{t})\label{eq:50-1}
	\end{equation}
	where $\beta\geq\Delta+\beta_{0}$ and $\beta_{0}>0$ with $\Delta\geq\Vert\frac{F_{d}}{m}+k_{s}\dot{\tilde{P}}\Vert_{\infty}$.
	Let Assumption \eqref{assum:Model assumpations} hold true. Then,
	the closed loop position error signals $\text{\ensuremath{\tilde{P}}}$
	is semi-global finite-time stable.
\end{thm}
\textbf{Proof}. Let us consider the following Lyapunov function candidate
$\mathcal{\mathcal{L}}:\mathbb{\mathbb{R}\times R}^{3}\times\mathbb{R}^{3}\rightarrow\mathbb{R}_{+}$
\begin{equation}
	\mathcal{\mathcal{L}}=k_{q}(1-\tilde{q}_{0})+\frac{1}{2}\Theta^{\top}J\Theta+\frac{1}{2}s_{t}^{\top}s_{t}\label{eq:Lyapunov_Total}
\end{equation}
with $s_{t}$ defined in \eqref{eq:s_x}-\eqref{eq:s_z} and $\Theta$
defined in \eqref{eq:auxiliary_variable}. The time derivative of
$\mathcal{\mathcal{L}}$ can be expressed as 
\begin{equation}
	\dot{\mathcal{\mathcal{\mathcal{\mathcal{L}}}}}=k_{q}\dot{\tilde{q}}_{0}+\Theta^{\top}J\dot{\Theta}+s_{t}^{\top}\dot{s}_{t}\label{eq:Lyapunov_Total_dot}
\end{equation}
In the light of \eqref{eq:quaterion_error}, one obtains $\dot{\tilde{q}}_{0}=\tilde{q}^{\top}\tilde{\Omega}.$
Recall the angular acceleration error dynamics in \eqref{eq:36},
by using the proposed control torque presented in \eqref{eq:control_troque-1},
one can rewrite \eqref{eq:36} as $J\dot{\tilde{\Omega}}=-k_{\theta}\Theta-k_{q}\tilde{q}+J\dot{\Psi}.$
From \eqref{eq:auxiliary_variable}, we have $J(\dot{\tilde{\Omega}}-\dot{\Psi})=-k_{\theta}\Theta-k_{q}\tilde{q}$,
thereby $J\dot{\Theta}=-k_{\theta}\Theta-k_{q}\tilde{q}.$ By using
the definition of translational acceleration $\dot{\tilde{V}}$ in
\eqref{eq:34} and by substituting the proposed virtual controller
in \eqref{eq:virtual_controller_main}, one can rewrite \eqref{eq:Lyapunov_Total_dot}
as follows:
\begin{flalign*}
	\dot{\mathcal{\mathcal{L}}} & =-k_{\theta}\Theta^{\top}\Theta+\left(k_{q}\Psi-\frac{-2\mathcal{T}}{m}R(Q)^{\top}[\bar{q}]_{\times}s_{t}\right)^{\top}\tilde{q}\\
	& -k_{t}s_{t}^{\top}\mathcal{S}(s_{t})+s_{t}^{\top}\frac{F_{d}}{m}+s_{t}^{\top}\varPsi+\beta_{t}s_{t}^{\top}k_{s_{t}}|\tilde{P}|^{\beta_{t}-1}\dot{\tilde{P}}
\end{flalign*}
The main role of $\Psi$ is to annihilate the attitude perturbation
in the translational error dynamics. By utilizing the proposed $\Psi$
in \eqref{eq:desgin_aux}, one finds
\begin{alignat}{1}
	\dot{\mathcal{\mathcal{L}}} & =-k_{\theta}\Theta^{\top}\Theta-k_{q}k_{\eta}\tilde{q}^{\top}\tilde{q}-k_{t}s_{t}^{\top}\mathcal{S}(s_{t})\nonumber \\
	& +s_{t}^{\top}\left(\underbrace{\frac{F_{d}}{m}+\beta_{t}k_{s_{t}}|\tilde{P}|^{\beta_{t}-1}\dot{\tilde{P}}}_{\Delta}\right)+s_{t}^{\top}\varPsi\label{eq:lap_modfied}
\end{alignat}
Suppose that $\Vert\frac{F_{d}}{m}+\beta_{t}k_{s_{t}}|\tilde{P}|^{\beta_{t}-1}\dot{\tilde{P}}\Vert_{\infty}\leq\Delta$,
one can rewrite \eqref{eq:lap_modfied} as follows
\[
\dot{\mathcal{\mathcal{L}}}\leq-k_{\theta}\Theta^{\top}\Theta-k_{q}k_{\eta}\tilde{q}^{\top}\tilde{q}-k_{t}s_{t}^{\top}\mathcal{S}(s_{t})+s_{t}^{\top}\Delta+s_{t}^{\top}\varPsi
\]
Consider $\varPsi=-\beta\mathcal{S}(s_{t})$, with $\beta\geq\Delta+\beta_{0}$
and $\beta_{0}>0.$ Consequently, one realize 
\begin{alignat}{1}
	\dot{\mathcal{\mathcal{L}}} & \leq-k_{\theta}\Theta^{\top}\Theta-k_{q}k_{\eta}\tilde{q}^{\top}\tilde{q}-\left(1+\beta_{0}\right)s_{t}^{\top}\mathcal{S}(s_{t})\label{eq:631}
\end{alignat}
Note that $s_{t}^{\top}\mathcal{S}(s_{t})\geq k_{s}||s_{t}||,$ where
$k_{s}\geq0$. Recall \eqref{eq:631}, we define the reaching time
$t_{r}$ such that $\mathcal{L}(t_{r})=0.$ By integrating \eqref{eq:631}
from both sides from $t=0$ to $t=t_{r}$, one finds
\begin{alignat*}{1}
	\mathcal{L}(0) & \geq t_{r}\left(k_{\theta}||\theta||^{2}+k_{q}k_{\eta}||\tilde{q}||^{2}+(1+\beta)k_{s}||s_{t}||\right)
\end{alignat*}
To find an upper bound on $t_{r}$, Let us consider the minimum values
of $||\theta||,$ \textbar\textbar$\tilde{q}||,$ and $||s_{t}||,$
thereby $t_{r}$ is given by
\[
t_{r}\leq\frac{\mathcal{\mathcal{L}}(0)}{\left(k_{\theta}||\theta||_{\text{min}}^{2}+k_{q}k_{\eta}||\tilde{q}||_{\text{min}}^{2}+(1+\beta)k_{s}||s_{t}||_{\min}\right)}
\]
For all $t>t_{r},$ $s_{t}=0$ such that the translational error dynamics
will be governed by $\dot{\tilde{p}}_{i}=k_{s_{t}}|\tilde{p}_{i}|^{\beta_{t}}\text{sgn}(\tilde{p}_{i})$
for $i\in\{x,y,z\}$ (visit \eqref{eq:s_x}-\eqref{eq:s_z}). It is
easy to get the closed form solutions of $\tilde{p}_{i}$(t) as follows:
\[
\tilde{p}_{i}(t)=\text{sgn}(\tilde{p}_{0i})\left(|\tilde{p}_{0i}|^{1-\beta_{t}}-k_{s_{t}}(1-\beta_{t})t\right)^{\frac{1}{1-\beta_{t}}}
\]
for all $t<t_{s_{i}},\tilde{p}_{0i}\neq0$ and
\[
\tilde{p}_{i}(t)=0,\hspace{1em}\text{for }(t\geq t_{s_{i}},\tilde{p}_{0i}\neq0)\text{ or }(t\geq0,\tilde{p}_{0i}=0)
\]
where $\tilde{p}_{0i}$ is the initial error and $t_{s_{i}}$ is the
settling time $\forall i=\{x,y,z\}.$ Thus, the maximum settling time
can be given by{\small{} $t_{s}=\text{max}\{\frac{1}{k_{s_{t}}\left(1-\beta_{t}\right)}|\tilde{p}_{0x}|^{1-\beta_{t}},\frac{1}{k_{s_{t}}\left(1-\beta_{t}\right)}|\tilde{p}_{0y}|^{1-\beta_{t}},\frac{1}{k_{s_{t}}\left(1-\beta_{t}\right)}|\tilde{p}_{0z}|^{1-\beta_{t}}\}.$
	To this end, one conclude that $\tilde{P}\rightarrow0,\forall t\geq t_{s}+t_{r}.$
}It is well-known that there are two physically identical equilibrium
for the unit-quaternion solutions $\tilde{q}_{0}=\pm1$. One equilibrium
solution is stable $(\tilde{q}_{0}=1)$, while the other is unstable
$(\tilde{q}_{0}=-1)$. Therefore, the best result that can be achieved
under the aforementioned topological obstruction is semi-global finite
time stabilization of error signals and this completes the proof.

\section{Numerical Results \label{sec: Numerical Results}}

\begin{figure*}[h]
	\centering\includegraphics[width=16.6cm,height=7.3cm]{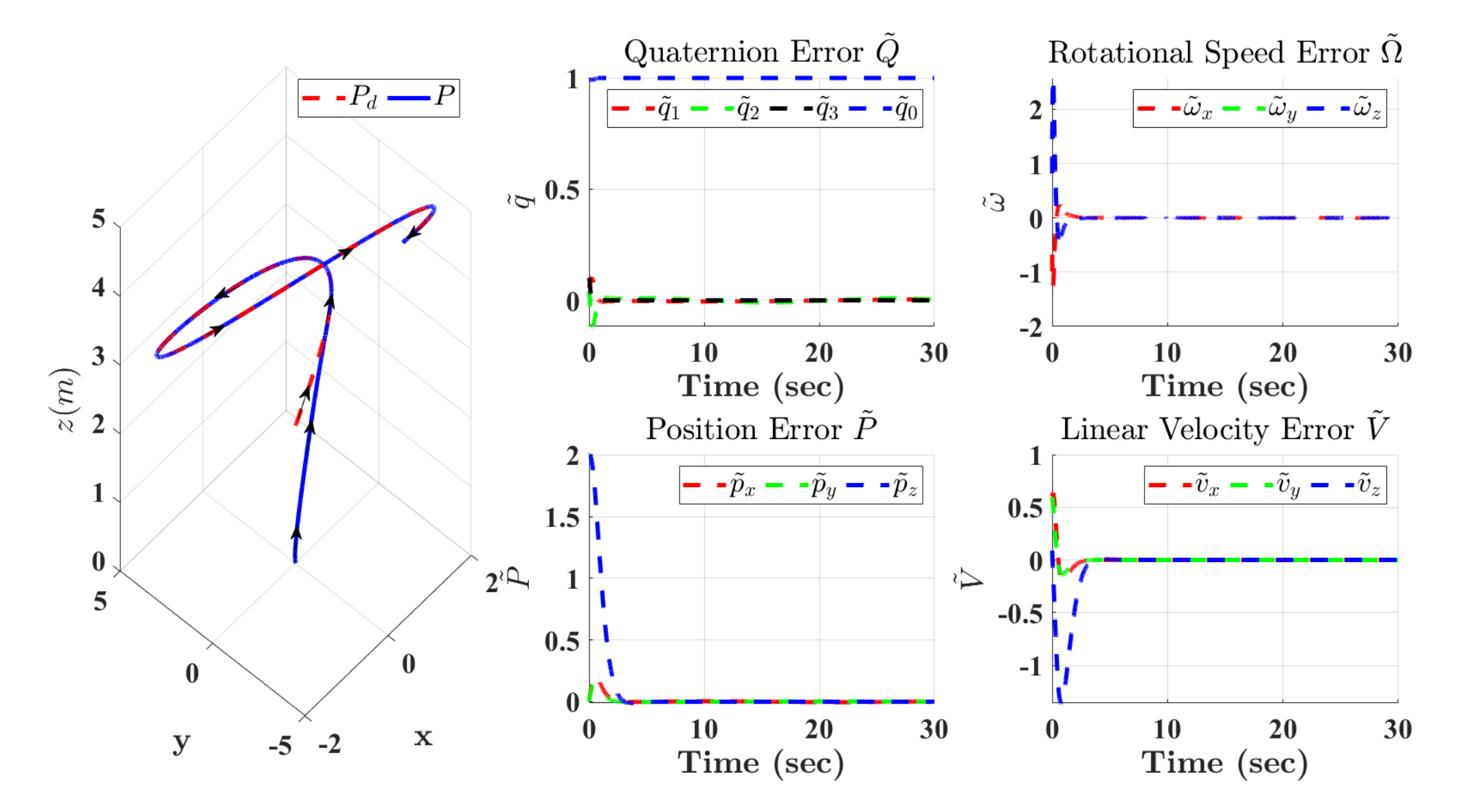}
	
	\caption{Output performance of the proposed control scheme. The left portion
		depicts the desired trajectory $P_{d}$ and the actual position of
		the quadrotor UAV. The right portion illustrates the convergence of
		error signals.}
	
	\label{fig_total_track}
\end{figure*}

In this section, the simulation results are presented to illustrate
the effectiveness of the proposed control scheme. The desired trajectory
is given by $P_{d}(t)=\left[4\sin(0.15t)\cos(0.15t),4\sin(0.15t),2+0.1t\right]^{\top}$
$\text{m}$. Note that $P_{d}(t)$ is selected to satisfy Assumption
\ref{A2.The-VTOL-UAV-current}.A2. The external disturbances is given
by $F_{d}(t)=\left[0.01\text{cos}(0.1t),0.01\text{cos}(0.1t),0.01\text{cos}(0.1t)\right]^{\top}$$\frac{\text{m}}{\text{s}^{2}}$.
Hence, we have $||\delta_{\text{f}}||=0.01$ (visit Assumption \ref{A2.The-VTOL-UAV-current}.A3).
The initial position, velocity, unit-quaternion, and rotational speed
are given, respectively, as $P(0)=[0,0,0]^{\top},$ $V(0)=[0,0,0]^{\top}$,
$Q(0)=[0.1699,0.3058,0.1699,0.9212]^{\top}$, and $\Omega(0)=[0,0,0]^{\top}$.
The total time is set to 30 seconds. The quadrotor UAV has a total
mass of $m=1\text{kg}$ and inertia of $J=\mathbb{I}_{3}$. The control
parameters used in the simulation are $k_{t}=2,$ $\beta=0.99,$ $k_{s_{t}}=5$,
and $k_{\eta}=k_{\theta}=k_{q}=15$. Fig. \eqref{fig_total_track}
demonstrates the output performance of the proposed control scheme
by comparing the trajectories of the desired and actual positions,
with all error signals converging within a finite time. The left portion
of Fig. \eqref{fig_total_track} shows the UAV starting with a significant
initialization error and smoothly converging to the desired trajectory.
The right part of Fig. \eqref{fig_total_track} highlights the error
convergence between the UAV and the desired trajectory: the quaternion
error $\tilde{q}=[\tilde{q}_{1},\tilde{q}_{2},\tilde{q}_{3}]^{\top}$,
angular velocity $\tilde{\Omega}$, position $\tilde{P}$, and linear
velocity $\tilde{V}$, all of which begin with large initialization
errors and converge to the origin. This underscores the robustness
and effectiveness of the proposed control strategy.

\section{Conclusion \label{sec: Conclusion}}

In this manuscript, a novel unified finite-time sliding mode control
approach for quadrotor Unmanned Aerial Vehicles (UAVs) is presented.
The proposed controller guarantees finite time convergence without
the need for time-scale separation. Specifically, it ensures the finite-time
stability of the position error while avoiding the traditional dependence
on the time-scale separation principle. To address singularity issues
in attitude representation, the controller is designed using quaternions.
The attitude controller acts as the primary control input, stabilizing
both translational and rotational dynamics through the introduction
of an auxiliary variable. The overall system stability is rigorously
analyzed and confirmed using Lyapunov stability. Finally, the effectiveness
of the control scheme is demonstrated through numerical simulations.

\balance
\bibliographystyle{IEEEtran}
\bibliography{ACC2025}
		
	\end{document}